\documentclass[12pt,preprint]{aastex}
\newcommand {\msun}{\mbox{M$_\odot$}}

\newcommand{\nc}{\newcommand}

\nc{\cratio}{\,$^{12}$C/$^{13}$C\,}
\nc{\teff}{$T_{\rm eff}$\,}
\nc{\logg}{log\,$g$\,}
\nc{\kms}{\,${\rm km\,s}^{-1}$\,}
\nc{\mic}{$\xi_{\rm t}$\,}
\nc{\feh}{\,${\rm [Fe/H]}$\,}

\shorttitle{Near-Infrared Spectroscopy of CEMP Stars}
\shortauthors{}

\begin{document}

\title{Near-Infrared Spectroscopy of Carbon-Enhanced Metal-Poor Stars. I.\\
A SOAR/OSIRIS Pilot Study}

\author{Timothy C. Beers\altaffilmark{1,2}, Thirupathi Sivarani, Brian Marsteller\altaffilmark{1}, YoungSun Lee\altaffilmark{1}}
\affil{Department of Physics \& Astronomy, CSCE: Center for the Study of Cosmic Evolution,
and JINA: Joint Institute for Nuclear Astrophysics, Michigan State University,
East Lansing, MI 48824, USA} \email{beers@pa.msu.edu,thirupathi@pa.msu.edu,marsteller@pa.msu.edu,leeyou25@msu.edu} 

\author{S. Rossi\altaffilmark{2}}
\affil{Instituto de Astronomia,  Geof\'{i}sica e Ci\^{e}ncias Atmosf\'{e}ricas, Departamento de Astronomia, 
Universidade de S\~{a}o Paulo, \\ 
Rua do Mat\~{a}o  1226, 05508-900 S\~{a}o Paulo, Brazil}
\email{rossi@astro.iag.usp.br}

\author{B. Plez}
\affil{GRAAL, Universit\'e de Montpellier II, F-34095 Montpellier Cedex 05,
France}
\email{Bertrand.Plez@graal.univ-montp2.fr}

\altaffiltext{1}{Visiting Astronomer, Kitt Peak National Observatory, 
which is operated by the Association of Universities
for Research in Astronomy, Inc. under cooperative agreement with the National
Science Foundation.}
\altaffiltext{2}{Visiting Astronomer, Cerro Tololo Interamerican Observatory,
which is operated by the Association of Universities
for Research in Astronomy, Inc. under cooperative agreement with the National
Science Foundation.}

\begin{abstract}

We report on an abundance analysis for a pilot study of seven Carbon-Enhanced
Metal-Poor (CEMP) stars, based on medium-resolution optical and near-infrared
spectroscopy. The optical spectra are used to estimate [Fe/H], [C/Fe], [N/Fe],
and [Ba/Fe] for our program stars. The near-infrared spectra, obtained during a
limited early science run with the new SOAR 4.1m telescope and the Ohio State
Infrared Imager and Spectrograph (OSIRIS), are used to obtain estimates of
[O/Fe] and $^{12}$C/$^{13}$C. The chemical abundances of CEMP stars are
of importance for understanding the origin of CNO in the early Galaxy, as well
as for placing constraints on the operation of the astrophysical s-process in
very low-metallicity Asymptotic Giant Branch (AGB) stars.

This pilot study includes a few stars with previously measured [Fe/H], [C/Fe],
[N/Fe],[O/Fe], $^{12}$C/$^{13}$C, and [Ba/Fe], based on high-resolution
optical spectra obtained with large-aperture telescopes. Our analysis
demonstrates that we are able to achieve reasonably accurate determinations of
these quantities for CEMP stars from moderate-resolution optical and
near-infrared spectra. This opens the pathway for the study of
significantly larger samples of CEMP stars in the near future. Furthermore, the
ability to measure [Ba/Fe] for (at least the cooler) CEMP stars should enable
one to separate stars that are likely to be associated with s-process
enhancements (the CEMP-s stars) from those that do not exhibit neutron-capture
enhancements (the CEMP-no stars).

\end{abstract}


\section{Introduction}

The large surveys for metal-poor stars conducted over the course of the past few
decades (e.g., the HK survey of Beers and colleagues; Beers, Preston, \&
Shectman 1985, 1992; Beers 1999), and the Hamburg/ESO survey of Christlieb and
collaborators (HES; Christlieb 2003) have revealed that a substantial fraction of
Very Metal-Poor (VMP; [Fe/H] $< -2.0$) stars exhibit enhanced ratios of carbon, 
[C/Fe] $> +1.0$ (this fraction is currently estimated to be at least 20\%, see
Lucatello et al. 2006). This frequency appears to rise with declining [Fe/H];
Beers \& Christlieb (2005) point out that 40\% of stars with [Fe/H] $< -3.5$,
based on high-resolution analyses, have [C/Fe] $>$ +1.0, including the two known
hyper metal-poor ([Fe/H] $< -$5.0) stars, HE~0107$-$5240 (Christlieb et al.
2002) and HE~1327$-$2326 (Frebel et al. 2005). These Carbon-Enhanced Metal-Poor
(CEMP) stars exhibit a wide variety of elemental abundance patterns (Beers \&
Christlieb 2005). The majority of CEMP stars ($\sim 80$\%) exhibit enhanced
s-process elements (Aoki et al. 2006a), and are referred to as CEMP-s stars.
Other CEMP stars exhibit strong enhancements of r-process elements (CEMP-r), or
the presence of enhanced neutron-capture elements associated with {\it both} the
r- and s-processes (CEMP-r/s). The class of CEMP-no stars comprises stars that,
in spite of their large C (and often N and O) overabundances with respect to Fe,
do not exhibit strong neutron-capture elements. Recently, CEMP stars have also
been found with large enhancements of the alpha elements (Norris, Ryan, \& Beers
2001; Aoki et al. 2002; Depagne et al. 2002), which Aoki et al. (2006b) refer to
as CEMP-$\alpha$ stars.
                                                                                                                             
The astrophysical sites of carbon production in CEMP stars are the subject of
much current observational and theoretical interest (Beers \& Christlieb 2005;
Ryan et al. 2005; Aoki et al. 2006a; Cohen et al. 2006; Johnson et al. 2006;
Jonsell et al. 2006; Karlsson 2006; Sivarani et al. 2006; Wanajo et al. 2006).
The carbon in the CEMP-s stars is very likely to have been produced by an
intermediate-mass Asymptotic Giant Branch (AGB) companion that has transferred
material to the presently observed star. The origin of the carbon in the other
CEMP classes is not yet fully understood. The CEMP-no stars are of special
interest, as it has been suggested that their C (as well as their N and O) may
have been produced by a primordial population of massive (20 $<$ ${\rm M}/\msun$
$<$ 100), rapidly-rotating, mega metal-poor ([Fe/H] $< -6$) stars, which are
predicted to have experienced significant mass loss (of CNO-enhanced material)
via strong winds (Hirschi et al 2006; Meynet et al 2006). Aoki et al. (2006a)
have shown that the CEMP-no stars are found preferentially at the lowest
metallicities (\feh $< -2.7$), while the CEMP-s stars are generally found in the
metallicity range $-2.7$ $\le$ \feh $\le$ $-$2.0. The CEMP-no stars exhibit
quite low \cratio ratios (in the range $4 \le$ \cratio $\le 10$), indicating
that a significant amount of mixing has occurred in their progenitor objects
(Aoki et al. 2006a; Sivarani et al. 2006). Recently, Piau et al. (2006) have
invoked processing by massive mega metal-poor progenitors in the early Galaxy to
account for the apparent absence of Li in the main-sequence turnoff hyper
metal-poor CEMP-no star HE~1327-2326. It is clear that the CEMP-no stars are of
fundamental importance for understanding the early evolution of elements in the
Galaxy.

In order to more fully test the association of CEMP-no stars with massive
primordial stars, and to better explore the nature of the s-process in
low-metallicity AGB stars, which is still rather poorly understood (Herwig
2005), we require measurements of the important elements C, N, and O for as
large a sample of CEMP stars as possible. C and N can be measured from CEMP
stars based on medium-resolution optical observations of molecular CH, C$_2$,
and CN features, while O remains a challenge, even at high spectral resolution.
Most previous high-resolution observations of CEMP stars have relied on
measurements of the O~I 7774\,\AA\ triplet lines, which are strongly affected by
NLTE effects (e.g., Asplund 2005). The most reliable oxygen abundances come from
the forbidden [OI] 6300\,\AA\ line, but this feature is quite weak at low
metallicities, and requires several hours of 8m-class telescope time per star in
order to obtain a detection. An attractive alternative is provided by
medium-resolution measurements of the strengths of the near-infrared first
overtone ro-vibrational bands of CO at 2.3$\micron$. Because the C abundances in
CEMP stars exceed the O abundances, essentially all of the O is locked up in CO
molecules; these lines thus provide a sensitive probe of the O abundance. In
addition, the large separation of the $^{13}$CO lines from the $^{12}$CO lines
at 2.3$\micron$ provides a straightforward means to measure the important mixing
diagnostic \cratio.
                                                                                                                                                                                                                                                 
In this paper we present the results of a pilot study, based on a combination of
medium-resolution optical and near-infrared spectroscopy, for a sample of seven
CEMP stars, including several with previously available results from
high-resolution spectroscopic studies. Similar optical spectra have already been
obtained for several hundred stars selected on the basis of their carbon
enhancement from the study of Christlieb et al. (2001). Roughly half of these
stars appear to be likely CEMP stars (Goswami 2005; Goswami et al. 2006;
Marsteller et al., in preparation). Near-infrared spectroscopy with SOAR/OSIRIS
is expected to be obtained for many of the confirmed CEMP stars from this sample
in the near future. 

Section 2 describes the spectroscopic observations and data reduction procedures
for our present study. In \S 3, we summarize the available photometry for our
program stars. Section 4 describes determinations of model atmosphere
parameters for our sample, and our techniques for deriving
estimates of abundances for C, N, O, Ba, and the \cratio ratio based on our
medium-resolution optical and near-infrared observations. Section 5 reports our
results for stars with previous analyses. In \S 6 we present a brief discussion
of these abundances in the context of previously observed CEMP stars. Plans for
future studies of CEMP stars at moderate spectral resolution, in the optical and
near-infrared, are described in \S 7.

\section{Observations and Data Reduction}

The medium-resolution optical spectra used in this study were obtained with the
GOLDCAM spectrograph on the KPNO 2.1m and with the RC Spectrographs on the KPNO
or CTIO 4m telescopes during the course of routine survey follow-up
observations of metal-poor candidates from the HES. In all cases the resolving
power of the spectra is $R = 3000$. The spectra cover the wavelength range from
3700 to 5000\,\AA , although the signal-to-noise ratios drop precipitously at
the blue end of the spectra (below 3800 \AA ). The optical spectra of our
program stars selected from the HES are shown are shown in Figure 1, along with
the best-fit models obtained from the synthetic spectra described below.

The near-infrared spectra were obtained with the new SOAR 4.1m telescope and the
OSIRIS spectrograph (DePoy et al. 1993) during the limited amount of observing
time set aside for early science programs with SOAR/OSIRIS during December 2005
and January 2006. The observing log for the near-infrared observations
(including coordinates of the stars observed, date of observation, exposure
times, and S/N ratio achieved) is provided in Table 1.  We used the
long slit and long camera (with focal ratio f/7), which provided a resolving
power $R = 3000$. Each star was observed on at least at two positions on the
slit in order to measure and adequately substract the sky background. We also
observed A0-type stars at the same airmass as the observations of the program
objects in order to correct for the presence of telluric lines in the spectra.
The total number of counts per pixel in each exposure were always kept well
below the known non-linearity regime of the detector. Corrections for remaining
small non-linearities in the detector counts, flat fielding, background
substraction, spectrum extraction, wavelength calibration, telluric line
subtraction, and continuum fitting were all accomplished using standard tasks
within IRAF (and CTIO extensions to this package)\footnote{IRAF is distributed
by the National Optical Astronomy Observatories, which is operated by the
Association of Universities for Research in Astronomy, Inc. under cooperative
agreement with the National Science Foundation.}.

Figure 2 shows the near-infrared spectra of the HES stars in our program, along
with the best-fit models obtained from the synthetic spectra described below. 
The near-infrared spectra for our other two program stars,
V~Ari and G~77-61, are discussed below.

\section{Optical and Near-Infrared Photometry}

Broadband $BVRI$ photometry for the HES stars is listed in Table 2, adopted in
part from Beers et al. (2006). For V~Ari and G~77-61, the $B-V$ photometry is
taken from the SIMBAD listing. Near-infrared $JHK$ photometry for all of our
program objects is available from the 2MASS Point Source Catalog (Skrutskie et
al. 2006). An estimate of the interstellar reddening along the line of sight to
each star is obtained from Schlegel, Finkbeiner, \& Davis (1998), which is also
listed in Table 2. We also make use of Table 6 from Schlegel et al. (1998) to
obtain the relative extinctions for various band passes.

\section{Analysis}
       
Below we describe the procedures employed in the analysis of our program stars.
Briefly, the measured optical and infrared colors are used to derive estimates
of the effective temperatures. The optical spectra are then used to estimate
metallicities. Based on these estimates of \teff\ and [Fe/H], theoretical
evolutionary tracks for stars with these parameters are used to
obtain estimates of surface gravity for each star. We then derive estimates
of the CNO, \cratio, and Ba abundance from the optical and near-infrared spectra.
Details are provided below.

\subsection{Estimated Stellar Atmospheric Parameters}

To obtain first-pass estimates of the effective temperatures for our program
stars, we employ the Alonso, Arribas, \& Mart{\'{\i}}nez-Roger (1996)
calibrations of \teff\ with various colors. The results are listed in Table 3.
The appropriate transformations between the different photometric systems
necessary for use of the Alonso et al. calibrations are carried out as described
in Sivarani et al. (2004). The $(B-V)$-based temperature estimates are quite low,
compared with those based on other colors, due to the strong effect of molecular
carbon absorption on the $B$-band flux. The $(V-K)$-based temperature is
expected to be superior, owing to the large leverage from the widely separated
wavelengths of the bands involved, and because both the $V$ and $K$ bands are
relatively free of potentially corrupting molecular carbon features. The $V-K$
colors from the synthetic carbon-enhanced models of Hill et al. (2000) also
agree very well with the empirical color-\teff relations obtained from the
Alonso calibrations. In the application of the Alonso calibrations, we have
adopted a metallicity of [Fe/H] $= -2.5$ for all of our program stars, with the
exceptions of HE~0322-1504 and HE~0534-4548, for which we assume \feh $= -2.0$.
Surface gravities, \logg, have been been estimated based on the Padova
evolutionary tracks for metallicities [Fe/H] $= -2.5$ and [Fe/H] $= -1.7$
(Girardi et al. 2000; Marigo et al. 2001). For reference, the next to last
column of Table 3 lists the atmospheric parameters obtained by previous analyses
based on high-resolution spectroscopy. The final column of Table 3 lists our
adopted atmospheric parameters. The initial [Fe/H] estimates, as described
below, are used for the selection of the appropriate model atmospheres.

Initial metallicities are estimated from the calibration of the variation of the
Ca~II KP index with de-reddened $(J-K)_0$ color described by Rossi et al.
(2005). These estimates are indicated as [Fe/H]$_{\rm I}$ in the second column of
Table 3. We did not have a medium-resolution spectrum of V-Ari, so we
have adopted the literature value of [Fe/H] (from Kipper \& Kipper 1990) in this
case. Also note that the $(J-K)_0$ color of G~77-61 lies just outside the color
region over which the Rossi et al. calibration is defined, so the value derived
as a first guess is somewhat uncertain. 

Refined estimates of metallicities are then estimated, based on fits of
synthetic spectra to the medium-resolution optical spectra. We begin with a fit
to the Ca~II K and H lines, and then carry out a cross-check by fitting the Ca~I
4227\,\AA\ line and the weak Fe~I feature at 4938\,\AA\ (which is a blend of the
Fe~I 4938.82\,\AA\ and 4939.69\,\AA\ lines). Most of the time the estimates
obtained in this manner agree well with one another. In the case of HE~0534-4548
we had to increase the value of \logg to consistently match the Ca~II K line and
the Ca~I 4227\,\AA\ lines. Thus, this simultaneous fit procedure provides an
additional check on the adopted surface gravity. The microturbulence velocity is
taken to be 2\kms, which is often assumed for such cool, C-rich stars. In any
case, we have no means of estimating a value for the microturbulence velocity
based on medium-resolution spectroscopic data; its effect on the derived
abundances is expected to be minimal, based on previous high-resolution work
(e.g., Aoki et al. 2006a). The adopted estimate of [Fe/H] is indicated as
[Fe/H]$_{\rm A}$ in the third column of Table 3. Note that our adopted
metallicity estimate is typically 0.4-0.6 dex more metal-rich than the initial
estimate obtained from the Rossi et al. (2005) calibration, indicating that this
calibration could still be improved upon.  

\subsection{Model Atmospheres and Adopted Linelists}

We use the NEWODF ATLAS9 models (with no overshoot; Castelli \& Kurucz 2003) as
a starting model for our syntheses, from which first-pass estimates of the CNO
are obtained. We next generate new models with the appropriate
CNO enhancement, using a version of ATLAS12 (Kurucz 1996) that runs
under Linux (ported to Linux from the original VAX version by Sbordone et al.
2004). For the synthesis we use atomic linelists mainly from the VALD database.
The CH and CN molecular linelist compiled by Plez (see Plez \& Cohen 2005) is
employed. The NH and C$_2$ molecular linelists are taken from the Kurucz
database (\texttt{http://kurucz.harvard.edu/linelists/linesmol/}). In the model
calculations and the abundance analysis we employ the solar abundance values
from Asplund, Grevesse, \& Sauval (2005). For the optical synthesis we use the
current version of the spectrum synthesis code \texttt{turbospectrum} (Alvarez
\& Plez 1998). For the K-band synthesis we use the SYNTHE code (Kurucz 1993).
We use the CH, CN, C$_{2}$ and CO linelists in the Kurucz database for the
synthesis of the near-infrared spectra.                                                                                                                             

\subsection{Determination of Abundance Ratios}

Carbon abundances for our program stars are derived from the C$_{2}$ swan bands
at 4736\,\AA . We did not make use of the CH features, since they are heavily
saturated in these cool stars. Spite et al. (2005) noted that one can obtain a
difference of about 0.2 dex in the derivation of C abundances between estimates
based on the CH and C$_{2}$ features. In Figure 3a we illustrate the sensitivity
of the optical spectra to variations in the adopted [C/Fe]. As can be seen in
this Figure, the optical C$_{2}$ features are extremely sensitive to relatively
small variations in the carbon abundance. Although the CN features also change
due to variations in adopted [C/Fe], we choose to employ the C$_{2}$ features
for deriving estimates of the carbon abundance. Note that the region surrounding
the CaII~K line is relatively insensitive to variations of [C/Fe], although for
extremely carbon-rich stars this still may have some effect on metallicity
estimates based on the strength of CaII~K (due to suppresion of the local
continuum; see Cohen et al. 2005). As a result, best results for metallicity
estimates of CEMP stars obtained from medium-resolution optical spectra should
be based on spectral synthesis calculations.

Nitrogen abundances are derived from the CN band located at 4215\,\AA; as a
cross-check we also attempt to match the CN bandhead at 3883\,\AA\ .
Unfortunately, the 3883\,\AA\ bandhead is heavily saturated at the very blue end
of our wavelength coverage, so this is only used as a consistency check. In
Figure 3b we illustrate the effect of variations in adopted [N/Fe] on the
optical spectra. The 4215\, \AA\ CN feature is sensitive to the N abundance,
however it is also quite sensitive to the C abundance. Our spectral coverage
does not extend blue enough to make use of the NH feature at 3360\,\AA\ for
estimation of the N abundance (e.g., see Johnson et al. 2006). The 3883\,\AA\ CN
feature bandhead is saturated, and the wings are just at the blue end of our
data, where we encounter problems with the fit of the continuum. For these
reasons, we only make use of the 4215\,\AA\ CN feature for estimation of the N
abundances for our program stars.

We estimate O abundances for our program stars from the near-infrared spectra,
based on the CO features at 2.3$\micron$. The C and N abundances derived from
the optical spectra are fixed in order to obtain our first-pass estimates of O
from the near-infrared spectra. Figure 3c illustrates the sensitivity of the CO
features in the near-infrared to variations in the O abundance. Once we estimate
the O abundances we revisit the C and N abundance estimates, since the C
abundance can, in principle, depend on the O abundance as a result of the
formation of CO molecules. Apart from the continuum absorption due to CN, there
also CN molecular line features around 2.3$\micron$, which provide additional
checks on the C and N abundances derived from the optical spectra. For V~Ari
(for which we do not have an optical spectrum available), as well as for G~77-61
(for which we could not obtain satisfactory fits to the C and N features), we
employ C and N abundances taken from the literature for estimation of the O
abundances. 

The isotope ratio \cratio\ can be derived from the optical spectra using the
$^{12}$C$_{2}$ bandhead at 4736\,\AA\ and the $^{12}$C$^{13}$C bandhead at
4745\,\AA. In many of our program stars the $^{12}$C$_{2}$ bandhead at 4736\,
\AA\ is saturated, and leads to a large derived \cratio. For this reason, we
prefer to adopt values for \cratio\ based on analysis of the near-infrared
spectra, when possible. In the near-infrared, we derive \cratio from the
$^{12}$CO and $^{13}$CO lines, which provide superior estimates of this ratio
than we are able to obtain from the optical spectra, as long as the quality of
the data is sufficient.

For the cool CEMP stars that comprise our pilot study, it is often possible to
detect the strong lines arising from Ba~II. Barium abundances are estimated from
the 4554\,\AA\ and 4934\,\AA\ Ba~II lines. The 4554\,\AA\ Ba~II lines are
blended with C$_{2}$ lines, while the 4934\,\AA\ lines are less affected by
C$_{2}$ bands. Examples of the Ba~II fits with synthetic spectra are shown in
Figure 4.

\subsection{Uncertainities in Derived Elemental Abundances and in \cratio}

The error in the derived C abundance depends on a number of factors, including
the uncertain placement of the pseudo-continuum, and the effects of errors in
estimated \teff, \logg, the microturbulence velocity $\xi$, as well as in the
adopted O abundance. Previous high-resolution spectroscopic studies (e.g., Hill
et al. 2000; Goswami et al. 2006) have indicated that the errors in derived 
\teff estimates based on $V-K$ are often less than 100~K.  If we simply adopt
a temperature error of 100~K, we find that variations at this level only
contribute to a change in the derived C abundance of 0.05 dex. The C$_{2}$
features in our optical spectra, which are our primary indicators of C
abundance, are very sensitive to errors in the surface gravity. As mentioned in
\S 4.1 above, we do not obtain our estimate of \logg{} from the spectra
themselves, but rather, from theoretical isochrones. Hence, one must be aware
that any systematic errors in these models will propogate into systematic errors
in our analysis. If we adopt an error for \logg{} of 0.5 dex, this corresponds
to changes in [C/Fe] by 0.2 dex. Microturbulence appears to have only a very
minor effect on the C$_{2}$ features of our moderate-resolution spectra. Changes
in $\xi$ of 1 \kms{} lead to only a 0.01 dex alteration in the derived C
abundance. As discussed below, our expected error in [O/Fe] determination is
0.22 dex. This error influences the derived error in [C/Fe], due to the
need to correct for CO formation. It is a relatively minor effect -- changes on
the O abundance by 0.2 dex result in changes in the C abundance by 0.05 dex. The
error in [C/Fe] arising from pseudo-continuum placement of the optical spectra
is about 0.2 dex. However, since we also make use of the near-infrared CN
features as a cross-check on the C and N abundances obtained from the optical
spectra, this effect should be minimized (the pseudo-continuum in the
near-infrared is much more evident). Taken as a whole, we expect that the above
errors give rise to a total error on [C/Fe] of $\sim$0.35 dex.

The derived [N/Fe] abundance depends on all the factors mentioned for [C/Fe], and
also depends on the [C/Fe] values. The total errors we estimate for [N/Fe] are
on the order of $\sim 0.45$ dex.

The derived [O/Fe] abundances can depend on errors in the determination of
\teff, \logg, $\xi$ and [C/Fe]. However, since all of our 
program stars are C-rich (C/O $>$ 1), the CO features are not sensitive to the
[C/Fe] values. The primary source of error arises from sensitivity of the CO
line strengths to \logg. We find that changes on the order of 0.5 dex in \logg{}
can give rise to changes in the derived [O/Fe] on the order of $\sim 0.15$ dex.
The total errors we estimate for [O/Fe] are on the order of $\sim$ 0.2 dex.

The \cratio{} ratio does not depend on any of the stellar parameters, however it
enters through saturation of the $^{12}$C$_{2}$ and $^{12}$CO features. The
$^{12}$CO features are rarely saturated for such metal-poor stars as are
included in our program. The primary source of error in the estimation of
\cratio{} from the near-infrared spectra comes from poor S/N. We estimate an
error in the \cratio{} of $\pm$2 at the typical S/N ratio of our spectra.

The errors in our derived Ba abundances that arise due to blending with C$_{2}$
features can be as large as 0.3 dex. The errors due to uncertainties in
\teff and \logg{} can lead to changes of about 0.2 dex in our estimated Ba
abundances. Hence, the total errors in derived Ba abundances based on our
medium-resolution spectra can be as large as 0.4-0.5 dex. This is still of
sufficient accuracy to at least differentiate between CEMP-s and CEMP-no stars,
as the CEMP-s stars typically exhibit [Ba/Fe] $ >$ +1.0.

We summarize the effects described above in Table 5. The source of each
uncertainty is listed in the first column, while its affect on the derived
elemental abundance or isotope ratio is listed in each of the remaining columns.
The total estimated uncertainty, obtained from the addition (in quadrature) of
each component of the error, is listed in the final row of this Table.

\section{Results for Stars with Previous Analyses}

\subsection{V~Ari}

V~Ari is a late-type classical CH star, with C and N enhancements, as well as
s-process enhancement, indicating that it likely underwent binary mass transfer
from an AGB companion (Kipper \& Kipper 1990; Sleivyte \& Bartkevicius 1990).
Van Eck et al. (2003) reported a large enhancement of Pb in this star, similar
to many other recently observed CEMP-s stars. Kipper \& Kipper (1990) derived
[C/Fe] = +2.1 and [N/Fe] = +1.5, based on fits to the violet C$_{2}$ and CN
features, respectively, adopting model atmospheres with parameters \teff =
3500\,K, \logg = 0.5, and \feh = $-2.5$. 

If we adopt the above atmospheric parameters and C and N abundances,
we derive an O abundance of [O/Fe] = +1.3 for V~Ari from our
near-infrared spectra. However, the fit we obtain is not very satisfactory. We
suspect that the C abundance derived by Kipper \& Kipper (1990) is too
high, since several of the synthetic CN features in the K-band region appear
too strong for their estimated C abundance. Note that Kipper \& Kipper
assumed a solar O abundance in order to derive their C,N abundances, a value
that is 20 times lower than our estimate of [O/Fe], assuming their other
parameters are correct. If we adopt their suggested C
abundance, we also obtain a very high value for the carbon isotope ratio,
\cratio $\sim$ 1000. Such high values of this ratio were also found by previous
investigations for V~Ari and TT~CVn (Aoki \& Tsuji 1997), a star thought to be
similar to V~Ari. However, CEMP stars reported in the literature that have
abundance estimates based on high-resolution spectroscopy exhibit a range of
\cratio{} between 4 and 100; most have a value around 10. Hence, we explored
changing the C and N abundances, rather than adopting the values given by Kipper
\& Kipper (1990), in order to better fit the near-infrared CN and CO features.
From this exercise, we obtain [C/Fe] = +1.5, [N/Fe] = +1.2, and [O/Fe] = +0.2.
We then obtain a \cratio = 90, which is consistent with some previously studied
CEMP stars. The high values for \cratio{} in the literature may be due to the fact
that the $^{12}$C$_{2}$ features are heavily saturated, which is often the case
for such low \teff stars. However, our derived \cratio{} for V~Ari is still
higher than that found in most CEMP stars, especially considering the fact that
the star is a giant, and one might expect lower values for \cratio{} due to
mixing with CN-processed material. One possible explanation could be that the surviving
companion of V~Ari has evolved into the giant stage, and gone through third
dredge-up. This would help to explain the high \cratio{}, which is similar to
other intrinsic AGB stars. However, Wannier et al.(1990) do not find any CO
J(1-0) emission or infrared excess, which is observed in most mass-losing AGB
stars. 

Figure 5a compares the observed near-infrared
spectrum of V~Ari with that of a synthetic spectrum generated from the Kipper \&
Kipper (1990) values for the C and N abundances. Clearly, the synthetic spectrum
exhibits substantially stronger CN than does the observed spectrum; the
continuum of the near-infrared synthetic spectrum in this region is strongly
depressed by the presence of the CN. Note that the CO bandhead strengths are not
as greatly affected. When we decrease the C abundance the synthetic spectrum
matches the observed spectrum much better. The CN features are weaker and the
continuum level is better matched. However, it is still the case that that the
observed and synthetic spectra do not match as well as for our other stars,
which could indicate that there remain significant errors in the estimated
stellar parameters for V~Ari, and/or in the models from which the synthetic
spectra were generated.

\subsection{G~77-61}

We initially adopted the C and N abundances for this star reported by Plez,
Cohen \& Melendez (2005), and obtained an estimate of the O abundance for
G~77-61 of [O/Fe] = +1.8, which is in fair agreement with their value ([O/Fe] =
+2.2), to within the expected errors. Plez et al. used higher resolution
near-infrared spectra than we have available ($R = 18000$), but they only fit a
single CO bandhead. We fit three bandheads of CO lines in the K-band (see Figure
5b), and also obtain a reasonably good fit for some of the CN features. It
should also be noted that somewhat different model atmospheres than ours were
adopted for their study. Our estimate of $^{12}$C/$^{13}$C = 5, based on the
optical spectrum for this star (the near-infrared spectrum of this star has
quite weak CO features, and is of insufficient S/N), agrees well with the ratio
obtained by Plez \& Cohen (2005). Our estimate of the Ba abundance for G~77-61,
[Ba/Fe] $< +0.5$, is consistent with the upper limit of [Ba/Fe] $< +1$ obtained
by these same authors.

\subsection{HE~1305+0007}

This star has been studied at high resolution in the optical by Goswami et al.
(2006). These authors employed atmospheric parameters \teff = 4750~K, \logg = 2.0
and [Fe/H] = $-$2.0. Their effective temperature and surface gravity estimates
were set based on the excitation balance and ionization equilibrium,
respectively, from on analysis of the detected Fe~I and Fe~II lines in their
spectrum. Our estimate of effective temperature, based on the $V-K$ color
(T$_{\rm eff}$ = 4560~K), is 200~K lower than theirs. At
least half of the discrepancy between our lower metallicity estimate for this
star, [Fe/H] = $-2.5$, can be accounted for by differences in the adopted
temperature. If we adopt the Goswami temperature estimate instead, we obtain a
metallicity estimate of [Fe/H] $= -2.2$. Goswami et al. (2006) found this star
to be enriched in Pb and other s-process elements. We estimate a Ba abundance
from the Ba~II 4934 \AA\ line of [Ba/Fe] = +2.9, which is 0.6 dex higher than
that found by Goswami et al. (2006), but is roughly consistent within the
expected errors of our procedure (see above). 

\section{Discussion}

The CNO abundance determinations, along with the \cratio{} and Ba abundances,
provide sufficient information to at least make a preliminary classification of
the stars in our pilot sample. It should be kept in mind that all of our program
objects, with the exception of G~77-61, are giants, hence they have almost
certainly undergone mixing processes that would have diluted material in the
outer envelope with gas that has undergone (at least) CN processing. This means
that the presently observed abundances of C and N, as well as the
\cratio{} ratio, may have been altered from their values prior to undergoing mixing. 
Any of the stars that acquired material as a result of mass transfer from an
AGB companion are also expected to have undergone some amount of
``evolutionary dilution'' (Lucatello et al. 2006), which would result in a net
reduction of the CNO and Ba abundances they received from their companion.
For the purpose of the following discussion, we have not included the (possibly
large) effects of NLTE or 3D atmospheric models on the derived abundances of the
light elements (see Asplund 2005).

\subsection{Nitrogen and Oxygen Abundances, and \cratio\ Measurements}
    
For our program stars, [C/Fe] always exceeds [N/Fe]. At present, the number of
stars known where [N/Fe] exceeds [C/Fe] is quite small (see the discussion by
Johnson et al. 2006). The [C/N] ratios of our program stars all fall at or below
+1.0; the lowest are V~Ari and HE~0322-1504, with [C/N] = 0.0 and +0.1,
respectively. These values are in the range that is often found for CEMP stars
(see, e.g., Figure 6 of Johnson et al. 2006). The models of AGB synthesis provided
by Herwig (2004) indicate that, if the origin of our program stars is associated
with mass transfer from very metal-poor AGB stars, these progenitors would be of
intermediate mass ($\sim$ 3-6 M$_\odot$).  
 
Only a small number of CEMP stars have measured O abundances reported in the
literature, owing to the difficulties of making this measurement, even at
high-resolution, in the optical spectra of such objects. This is particularly
true for the cooler stars we have considered in our pilot sample, as the low
temperatures enhance the strength of the molecular bands associated with carbon.
Thus, it is difficult to put our present measurements into a larger context;
this will have to wait for the assembly of a sufficiently large sample of CEMP
stars with available O measurements in the near future. The measurements we have
carried out from our near-infrared spectra demonstrate that [O/Fe] is strongly
correlated with [C/Fe]; see Figure 6. This behavior is consistent with that
expected if the envelope material has been accreted from an AGB companion that
has undergone third dredge-up mixing. The abundances range from a low value of
[O/Fe] = +0.1 (HE~0534-4548) to a high of [O/Fe] = +1.8 (G~77-61). These values
all fall in the range which Sivarani et al. (2006) associate with non Hot Bottom
Burning (non-HBB), low mass (0.8-3.0 \msun), VMP AGB stars.

The carbon isotope measurements, with the exception of the large value for V~Ari
(\cratio\ = 90) and the intermediate value for HE~1045-1434 (\cratio\ = 20), are
all quite low, \cratio\ $< 10$. Such values are often found in CEMP stars of
various classes (see the compilation in Table 4 of Sivarani et al. 2006), and
may be driven primarily by the extent that mixing processes have altered the
observed surface isotopic ratio. The higher measured [C/Fe] and the intermediate
\cratio{} for HE~1045-1434, as compared to the other stars studied here, indicates
that less dilution and CN processing has occured during its giant-branch phase.

\subsection{The CEMP-s Stars}
       
Five of the seven stars (the exceptions being G~77-61 and HE~0534-4548) have
large [Ba/Fe] ratios, consistent with their identification as CEMP-s stars. Aoki
et al. (2006a) report that 80\% of their large sample of CEMP stars can be
classified as CEMP-s, and are likely to be associated with AGB mass transfer to
the presently observed companion star. The CNO abundances of these stars,
discussed above, are consistent with origin in intermediate-mass non-HBB AGB
stars. The CEMP-s stars include, as a subclass, the so-called lead (Pb) stars.
From the high-resolution study of HE~1305+0007 by Goswami et al. (2006), we are
aware that this star from our program is a member of this class. High-resolution
spectroscopic observations of the rest of the stars in our pilot sample will be
required in order to check if there are additional members of this class in our
sample. HE~1305+0007 has also been shown by Goswami et al. (2006) to be a
CEMP-r/s star, a group of stars that exhibit large enhancements of elements
associated with the r-process, in addition to s-process enhancements (see Beers
\& Christlieb 2005, and references therein).

\subsection{The CEMP-no Stars}

G~77-61 and HE~0534-4548 exhibit significantly lower [Ba/Fe] than found for the
CEMP-s stars in our program. We only have an upper limit on the [Ba/Fe] for
G~77-61, so it remains possible that it satisfies the Beers \& Christlieb (2005)
definition for CEMP-no stars ([Ba/Fe] $< 0$), while HE~0534-4548 might be
considered a Ba-mild star, since its [Ba/Fe] is above solar ([Ba/Fe] $\sim
+0.6$), but clearly lower than the [Ba/Fe] ratios of the CEMP-s stars in our
sample. The [Ba/Fe] ratio for this star should be confirmed by additional
high-resolution spectroscopic observations. As mentioned in the introduction, the CEMP-no
stars are of particular interest, since they may be associated with massive mega
metal-poor progenitors, rather than AGB stars.  

\section{Future Studies}

This paper has presented an analysis of medium-resolution optical and
near-infrared spectroscopy for a small number of CEMP stars. We have
demonstrated that this approach is capable of obtaining measurements of the
critical elemental abundances needed to discriminate CEMP-s stars from CEMP-no
stars in an efficient manner. It is our intention to obtain a greatly
expanded set of near-infrared spectroscopy with the SOAR 4.1m telescope for on
the order of 100 CEMP stars, including a larger number of stars with previous
high-resolution data available, in the near future. Optical medium-resolution
spectroscopy for this sample is already available. The availability of [Fe/H],
[C/Fe], [N/Fe], [O/Fe], \cratio\ , and [Ba/Fe] for such a large sample will
provide data (in particular for O and the \cratio ) that complements available
high-resolution studies, and will also identify stars of particular interest for
future inspection at high spectral resolution. 

\acknowledgments

The authors express gratitude to Steve Heathcote and Bob Blum for several useful observational
tips, and for assistance with OSIRIS during SOAR early science observations. We
are also greatful for the capable handling of the telescope by Patricio Ugarte and
Alberto Alvarez. The authors also wish to thank an anonymous referee for comments
that greatly improved the final manuscript. T.C.B., T.S., B.M., and Y.L.
acknowledge partial funding for this work from grant AST 04-06784, as well as
from grant PHY 02-16783: Physics Frontiers Center/Joint Institute for Nuclear
Astrophysics (JINA), both awarded by the U.S. National Science Foundation. S.R.
acknowledges partial support for this work from CNPq, FAPESP, and Capes.

\begin{deluxetable}{lrrrrr}
\tablecolumns{6}
\tablewidth{0pt}
\tablenum{1}
\tablecaption{Observing Log of the Near-Infrared Spectroscopy}
\tablehead{
\colhead{Star} & \multicolumn{2}{c}{Coordinates}  & \colhead{Date}  &  \colhead{Exposure} &  \colhead{S/N}\\
\colhead{}     & \colhead{RA (2000)} & \colhead{ Dec (2000)} & \colhead{y/m/d}    &  \colhead{(sec)}    &\colhead{at 2.3$\mu$m}
}
\startdata
V~Ari           &  02 15 00.0 &   +12 14 23.6 & 2005-12-28   &  600 x 2   &  40  \\
G~77-61         &  03 32 38.3 &   +01 57 57.9 & 2005-12-27   & 2400 x 2   &  55  \\
HE~0322-1504    &  03 24 40.1 & $-$14 54 24.0 & 2005-12-28   & 1800 x 2   &  30   \\
HE~0507-1430    &  05 10 07.6 & $-$14 26 32.0 & 2005-12-28   & 2400 x 2   &  55   \\
HE~0534-4548    &  05 36 06.1 & $-$45 46 56.0 & 2005-12-27   & 1800 x 2   &  45   \\
HE~1045-1434    &  10 47 44.1 & $-$14 50 23.0 & 2005-12-28   & 1800 x 2   &  30   \\
HE~1305+0007    &  13 08 03.8 & $-$00 08 48.0 & 2006-01-18   &  900 x 2   &  60   \\
\enddata

\end{deluxetable}

\begin{deluxetable}{lrrrrrrrr}
\tablecolumns{9}
\tablewidth{0pt}
\tablenum{2}
\tablecaption{Optical and Near-Infrared Photometry}
\tablehead{
\colhead{Star}   &  \colhead{$V$}   & \colhead{$E(B-V)$} & \colhead{$B-V$}  &  \colhead{$V-R$}  &
\colhead{ $V-I$}   &  \colhead{$J$}  & \colhead{$H$}  & \colhead{$K$}
}
\startdata
V~Ari        &   8.45   &  0.142 & 2.05  & \nodata & \nodata &   5.104 &  4.620 &  4.364 \\
G~77-61      &  13.97   &  0.109 & 1.73  & \nodata & \nodata &  11.470 & 10.844 & 10.480  \\
HE~0322-1504 &  14.177  &  0.056 & 1.468 & 0.667   &  1.244  &  12.105 & 11.533 & 11.340  \\
HE~0507-1430 &  14.486  &  0.121 & 1.541 & 0.707   &  1.296  &  12.325 & 11.717 & 11.575  \\
HE~0534-4548 &  14.034  &  0.052 & 1.477 & 0.660   &  1.285  &  11.741 & 11.129 & 10.926  \\
HE~1045-1434 &  14.639  &  0.075 & 1.454 & 0.564   &  1.000  &  12.935 & 12.449 & 12.244  \\
HE~1305+0007 &  12.223  &  0.022 & 1.459 & 0.682   &  1.152  &  10.247 &  9.753 &  9.600  \\
\enddata
\end{deluxetable}

\begin{deluxetable}{lccccccccccc}
\rotate
\tabletypesize{\scriptsize}
\tablecaption{Stellar Parameters}
\tablewidth{0pt}
\tablenum{3}
\tablehead{
\colhead{Object} & \colhead{[Fe/H]$_{\rm I}$} & \colhead{[Fe/H]$_{\rm A}$}   & \colhead{T$(B-V)$} & \colhead{T$(V-R)$} &
\colhead{T$(V-I)$} & \colhead{T$(R-I)$} & \colhead{ T$(V-K)$} & \colhead{T$(J-H)$} & \colhead{ T$(J-K)$}  & \multicolumn{2}{c}{Atmospheric Parameters}\\
\colhead{} & \colhead{} & \colhead{} &\colhead{} &\colhead{} &\colhead{} &\colhead{} &\colhead{} &\colhead{} &\colhead{} &\colhead{High$-$Res} &\colhead{Adopted}
}
\startdata
V~Ari          & $-$2.5 & $-2.5$ & 2685 &\nodata&\nodata&\nodata& 3865 & 4905 & 4561 &3500,0.5,$-$2.5& 3500,0.5,$-2.5$  \\
G~77-61        & $-$4.6 & $-5.2$ & 2833 &\nodata&\nodata&\nodata& 4124 & 4271 & 3934 &4000,5.1,$-$4.0& 4000,5.1,$-4.0$  \\
HE~0322-1504   & $-$2.4 & $-2.0$ & 3143 &4334   &4903   &4535   & 4462 & 4458 & 4363 & \nodata       & 4460,0.8,$-2.0$  \\
HE~0507-1430   & $-$3.0 & $-2.4$ & 3135 &4339   &4957   &4710   & 4562 & 4414 & 4500 & \nodata       & 4560,1.2,$-2.4$  \\
HE~0534-4548   & $-$2.2 & $-1.8$ & 3129 &4347   &4841   &4318   & 4251 & 4322 & 4236 & \nodata       & 4250,1.5$\tablenotemark{a}$,$-$1.8  \\
HE~1045-1434   & $-$2.9 & $-2.5$ & 3180 &4726   &5383   &5484   & 4947 & 4800 & 4593 & \nodata       & 4950,1.8,$-$2.5  \\
HE~1305+0007   & $-$2.8 & $-2.5$ & 3116 &4230   &4958   &4972   & 4558 & 4696 & 4635 &4750,2.0,$-$2.0& 4560,1.0,$-$2.5  \\
\enddata

\tablenotetext{a} {The \logg{} derived from the isochrones did not match with Ca~I 4224\,\AA\ and the CaII~K synthesis.
The final adopted value is based on spectrum synthesis.}

\end{deluxetable}

\begin{deluxetable}{lrrrrrr}
\tablecaption{Elemental Abundances}
\tablewidth{0pt}
\tablenum{4}
\tablehead{
\colhead{Star}   & \colhead{[Fe/H]} &  \colhead{[C/Fe]}   &  \colhead{[N/Fe]}  &
\colhead{[O/Fe]}  &  \colhead{ \cratio}  & \colhead{[Ba/Fe]}
}
\startdata
V~Ari\tablenotemark{a}&  $-$2.5   &+1.5      &  +1.5     &  +0.2     &   90$\pm$10& \nodata\tablenotemark{b}  \\
G~77-61\tablenotemark{c}&$-$4.0   &+3.2      &  +2.2     &  +1.8     &   5$\pm$2\tablenotemark{d} &  $< +0.5$   \\
HE~0322-1504          &  $-$2.0   &+2.3      &  +2.2     &  +1.0     &    6$\pm$2 & +2.8     \\
HE~0507-1430          &  $-$2.4   &+2.6      &  +1.7     &  +1.1     &    9$\pm$2 & $\sim$ +1.3     \\
HE~0534-4548          &  $-$1.8   &+1.5      &  +1.1     &  +0.1     &    5$\pm$1 & $\sim$ +0.6     \\
HE~1045-1434          &  $-$2.5   &+3.2      &  +2.8     &  +1.8     &   20$\pm$2&  +3.0     \\
HE~1305+0007          &  $-$2.5   &+2.4      &  +1.9     &  +0.8     &   9$\pm$2 & +2.9     \\
\enddata
\tablenotetext{a}{The CNO abundances are based on only the CN features and the CO
features in the near-infrared spectra.}
\tablenotetext{b}{This star is known to be s-process rich (Van Eck et al. 2003).}
\tablenotetext{c}{[C/Fe], [N/Fe] values are taken from Plez, Cohen \& Melendez
(2005).}
\tablenotetext{d}{The \cratio{} is estimated from the 4737 \AA{} $^{12}$C$_{2}$
and the 4740 \AA{} $^{13}$C,$^{12}$C lines in the optical spectrum.}
\end{deluxetable}

\begin{deluxetable}{lrrrrr}
\tablecaption{Uncertainities in Derived Abundances}
\tablewidth{0pt}
\tablenum{5}
\tablehead{
\colhead{Uncertainities} & \colhead{$\sigma$([C/Fe])}   & \colhead{$\sigma$([N/Fe])} &  \colhead{$\sigma$([O/Fe])}   &  \colhead{$\sigma$(\cratio)}  &
\colhead{[Ba/Fe]}  
}
\startdata
$\sigma$(\teff) (100~K)  &  0.05     & 0.05  &  0.10 & \nodata & 0.2 \\ 
$\sigma$(\logg) (0.5 dex)&  0.20     & 0.20  &  0.15 & \nodata & 0.2 \\
$\sigma$(\mic) (1\kms)   &  0.01     & 0.01  &  0.01 & \nodata & 0.1 \\
$\sigma$([C/Fe])         & \nodata   & 0.35  &\nodata& 1.0     &\nodata \\
$\sigma$([O/Fe])         &  0.22     &\nodata&\nodata&\nodata  &\nodata\\
$\sigma$(cont)\tablenotemark{a}      &  0.20 & 0.20  &\nodata  &2.0      &0.4 \\
$\sigma$(tot)            &  0.35     & 0.46  & 0.22  &2.2      &0.5 \\ 
\enddata
\tablenotetext{a}{Error due to pseudo-continuum placement.}
\end{deluxetable}

\clearpage
\begin{figure}
\figurenum{1}
\epsscale{0.80}
\plotone{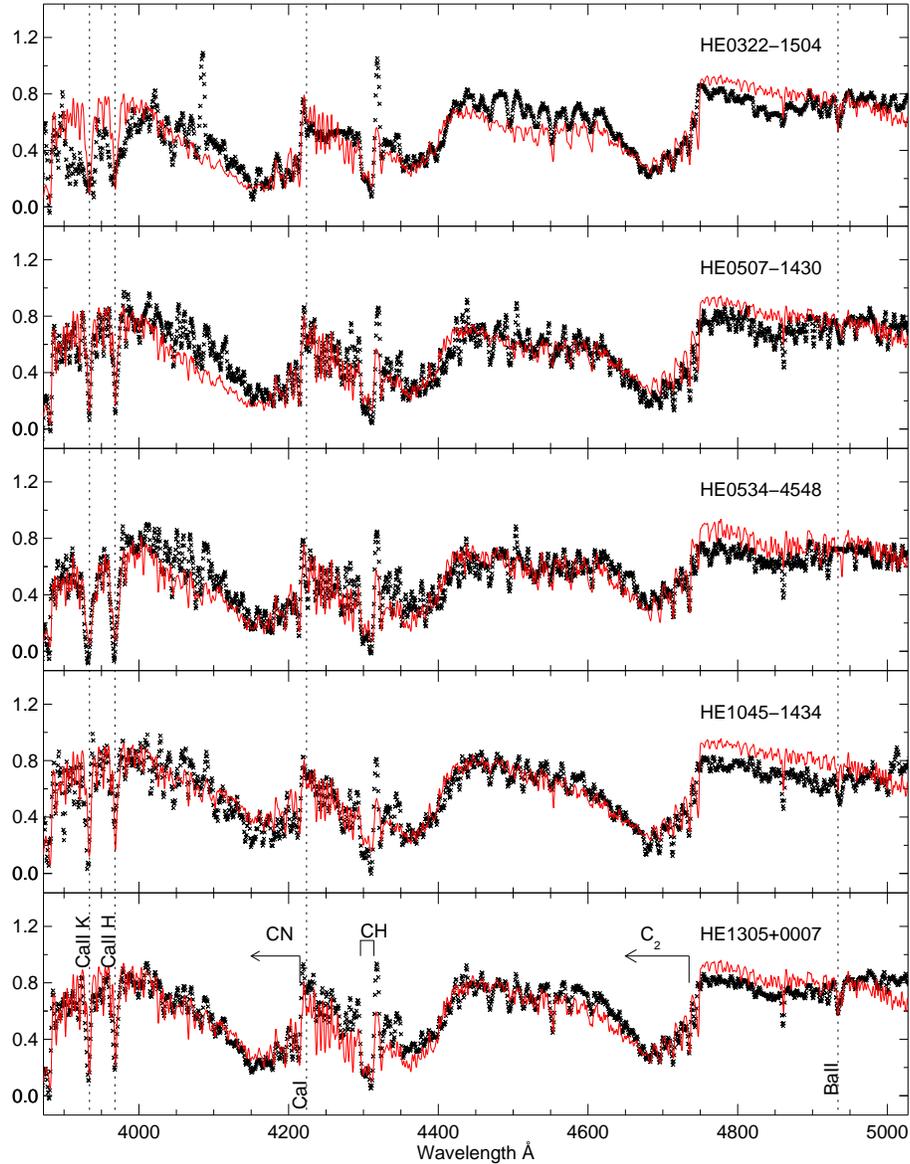}
\caption{Medium-resolution optical spectra of the HES program stars,
obtained with the KPNO 2.1m and/or the KPNO/CTIO 4m
telescopes. The crosses are the data, while the solid lines are the best-fit
synthetic spectra based on our adopted atmospheric parameters and CNO
abundances. The prominent absorption features used in our analysis are labelled
on the spectrum of HE~1305+0007. }
\end{figure}
\clearpage

\begin{figure}
\figurenum{2}
\epsscale{0.80}
\plotone{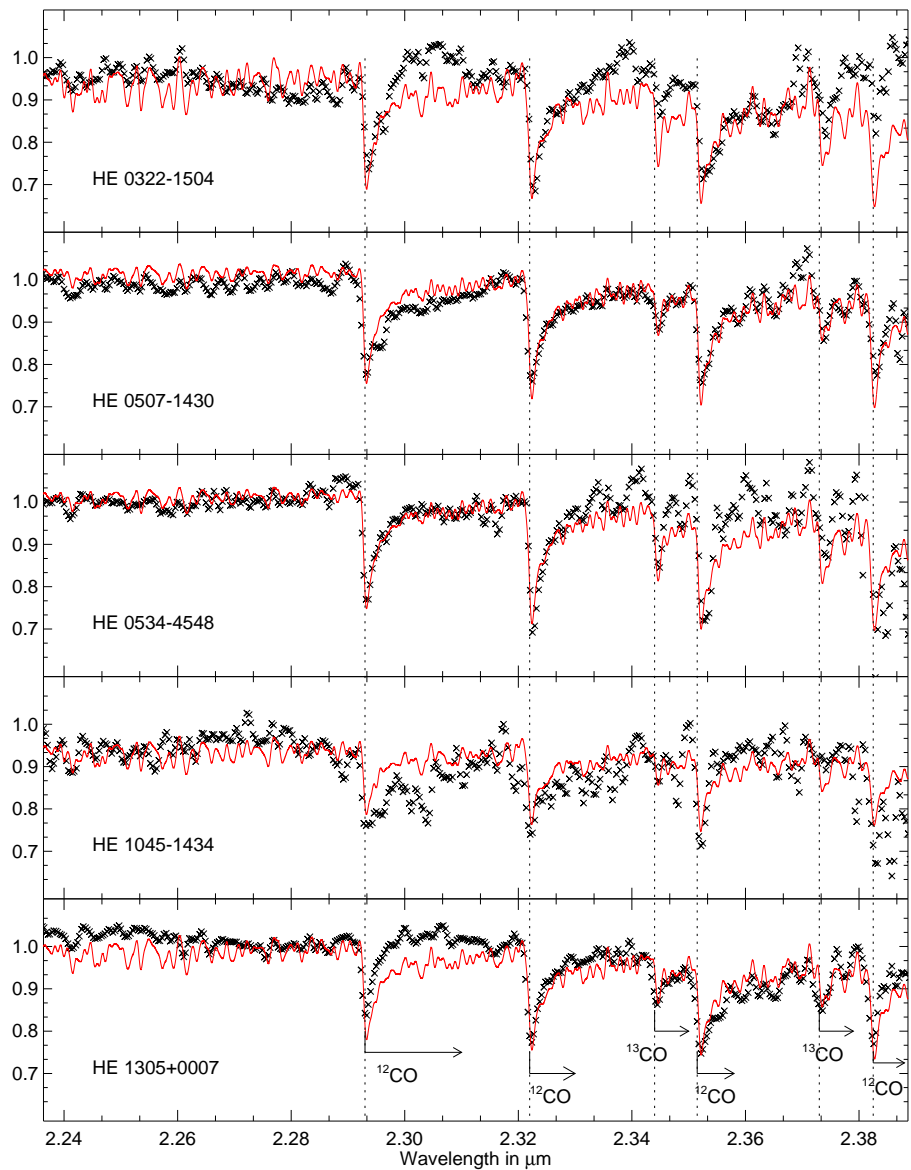}
\caption{Medium-resolution near-infrared K-band spectra of the HES program stars,
obtained with OSIRIS on the SOAR 4.1m telescope. The crosses are the data, while
the solid lines are the best-fit synthetic spectra based on our adopted
atmospheric parameters and CNO abundances. The prominent absorption features
used in our analysis are labelled on the spectrum of HE~1305+0007.}
\end{figure}
\clearpage

\begin{figure}
\figurenum{3}
\epsscale{0.7}
\plotone{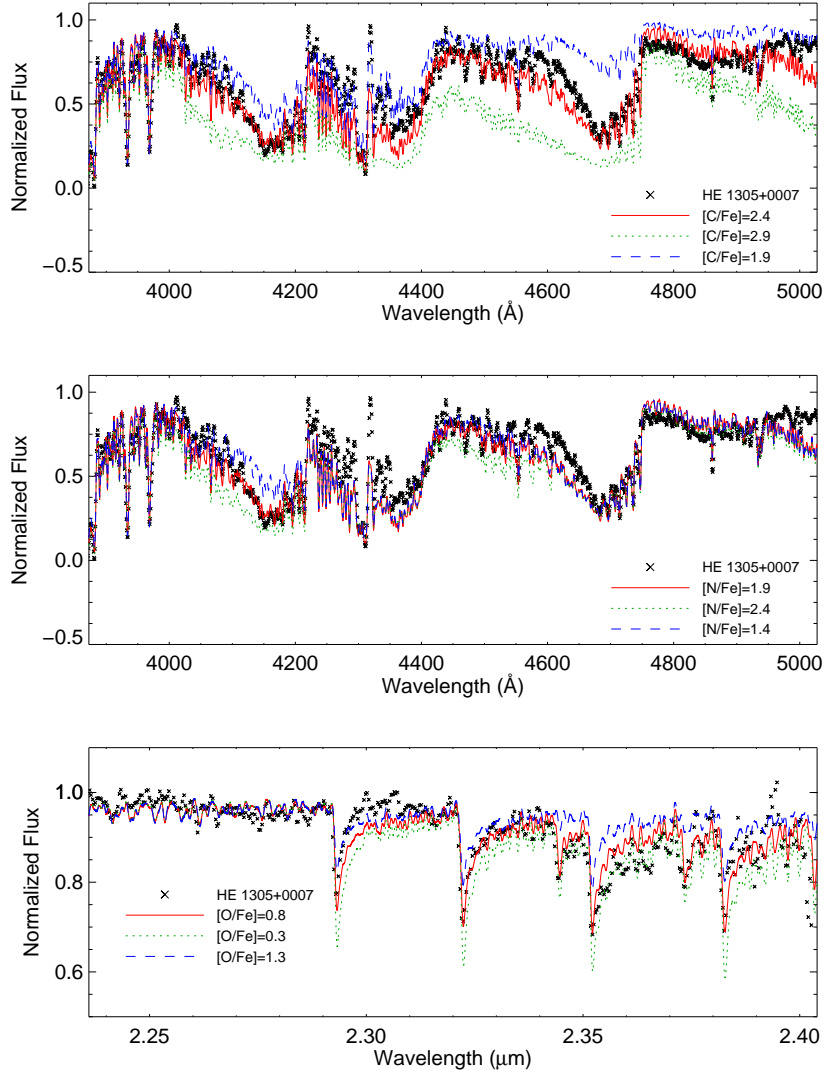}
\caption{\small{(a) (Upper Panel) The observed optical spectrum of HE~1305+0007 is compared with
synthetic spectra with C abundances taken to be 0.5 dex higher (dotted line) and 0.5
dex lower (dashed line) than our adopted C abundance ([C/Fe] = +2.4).
As can be seen, the C$_{2}$ feature is very sensitive to small changes in
[C/Fe] for a fixed N abundance; the CN feature exhibits somewhat lower
sensitivity. (b) (Middle Panel)  The observed optical spectrum of HE~1305+0007 is compared with
synthetic spectra with N abundances taken to be 0.5 dex higher (dotted line) and 0.5
dex lower (dashed line) than our adopted N abundance ([N/Fe] = +1.9). The CN
feature clearly changes in response to the different N abundance, but by less than
the response to similar changes in the C abundance for a fixed N abundance.
(c) (Lower Panel) The observed near-infrared (K-band) spectrum of HE~1305+0007 is compared with
synthetic spectra with O abundances taken to be 0.5 dex higher (dotted line) and 0.5
dex lower (dashed line) than our adopted O abundance ([O/Fe] = +0.8).
The CO features are seen to be quite sensitive to changes in the O abundance for
fixed C and N abundances.}}
\end{figure}
\clearpage

\begin{figure}
\figurenum{4}
\epsscale{1.00}
\plottwo{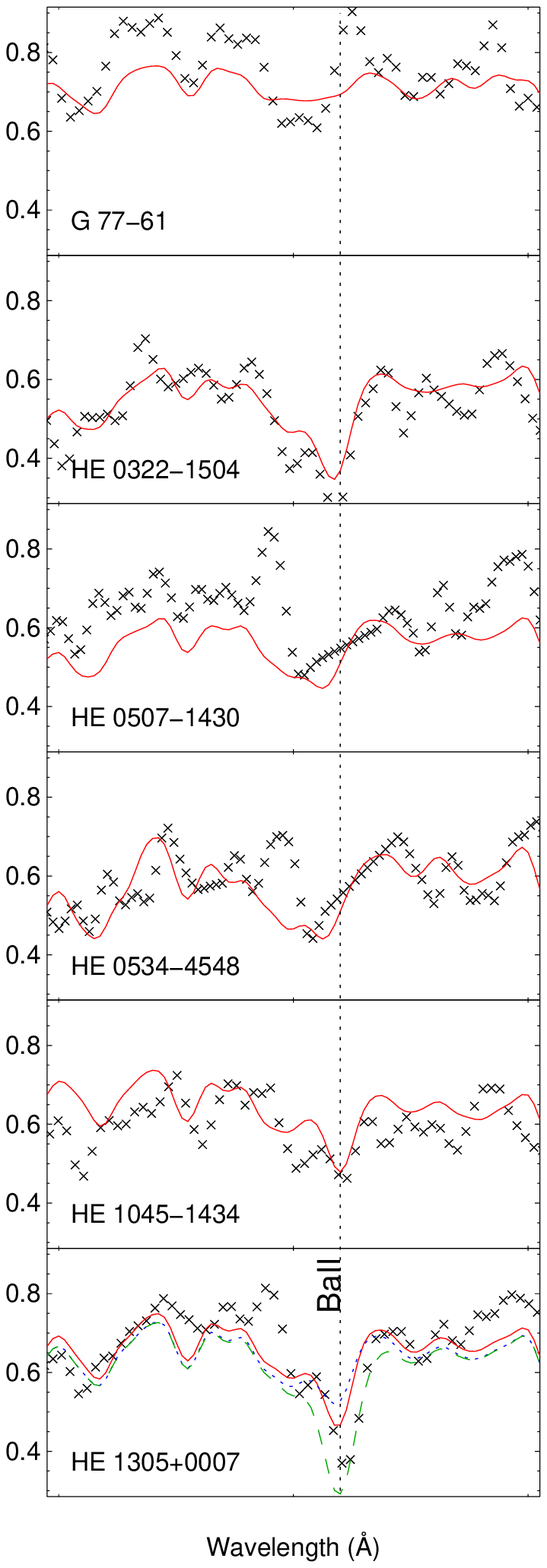}{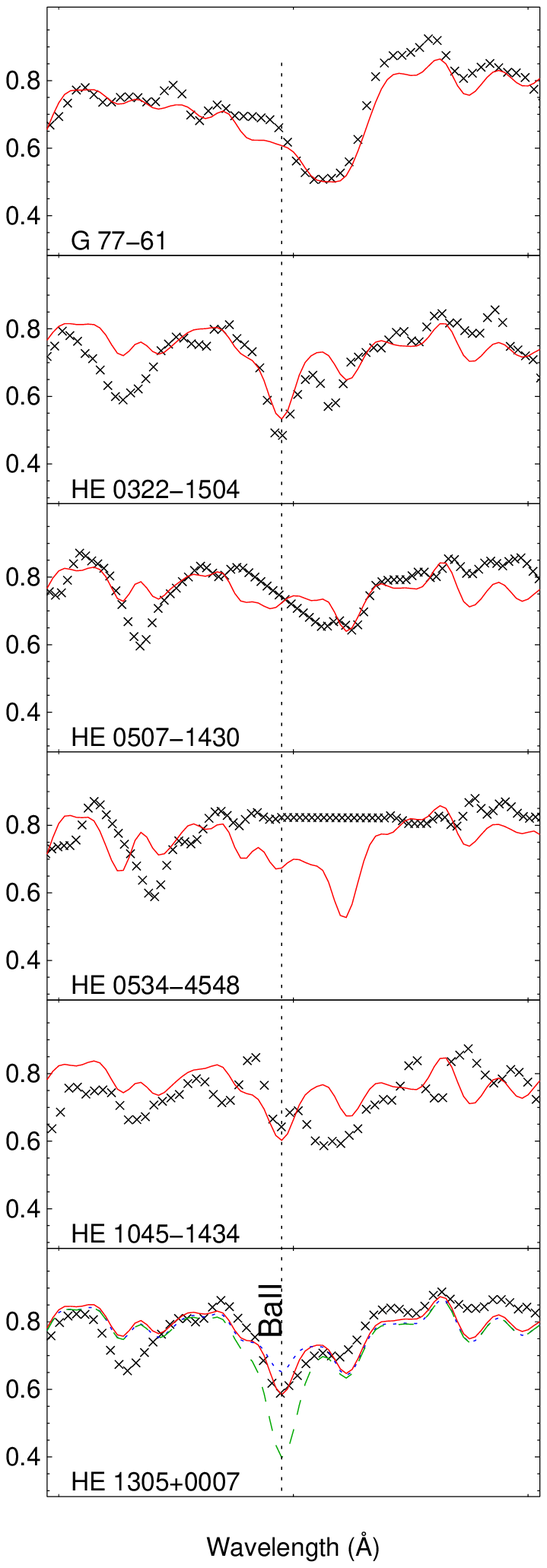}
\caption{Observed (crosses) and best-fit synthetic (solid lines) optical spectra for the 
Ba~II 4934 \AA\ and BaII 4554 \AA\ lines. HE~0322-1504, HE~1045-1434, and
HE~1305+0007 exhibit clearly detectable Ba~II lines; for the other stars the
fits are less certain. The results listed in Table 3 rely mostly on the Ba~II
4934 \AA\ lines, with the exception of HE~0534-4548. The dotted, solid, and
dashed lines plotted for HE~1305+0007 are synthetic spectra for Ba
abundances [Ba/Fe] = +1.9, +2.9, and +3.9, respectively. }
\end{figure}
\clearpage

\begin{figure}
\figurenum{5}
\epsscale{1.00}
\plotone{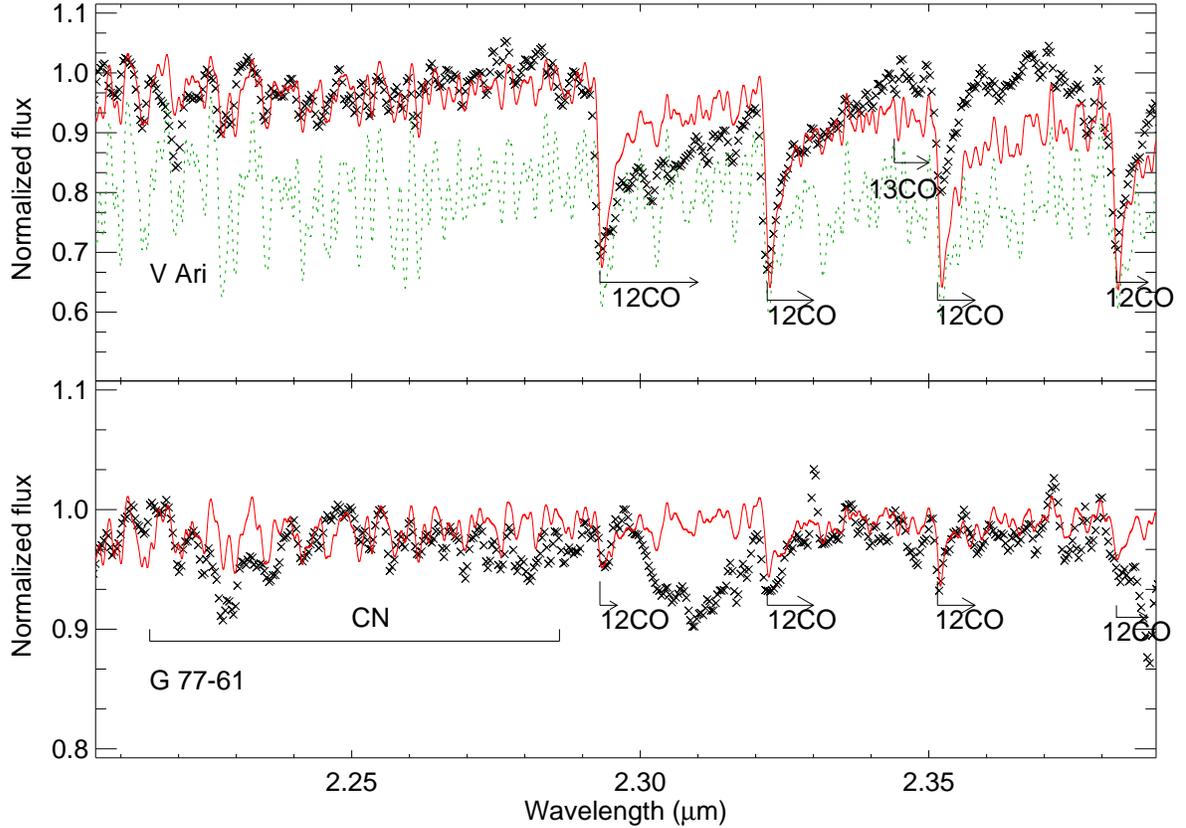}
\caption{(a) (Upper Panel) The solid line indicates the best fit of the observed
near-infrared (K-band) spectrum (crosses) for V~Ari, obtained using our adopted
atmospheric parameters and the CNO abundances
listed in Table 4. The dotted line shows the synthetic spectrum obtained from
adoption of the C and N abundances of Kipper \& Kipper (1992), and the O
abundance listed in Table 4.  Note the extremely strong CN feature
that results in this case; it clearly does not match the observed spectrum.
(b) (Lower Panel) The solid line indicates the best fit of the near-infrared
spectrum (crosses) for G~77-61, based on our adopted atmospheric parameters and CNO
abundances.  The CO feature at 2.3 $\micron$, which is not well fit by the model,
is affected by the presence of telluric lines that could not be removed
adequately from the spectrum.} 
\end{figure}    
\clearpage
    
\begin{figure}
\figurenum{6}
\rotatebox{90}{\resizebox{14.5cm}{!}{\includegraphics{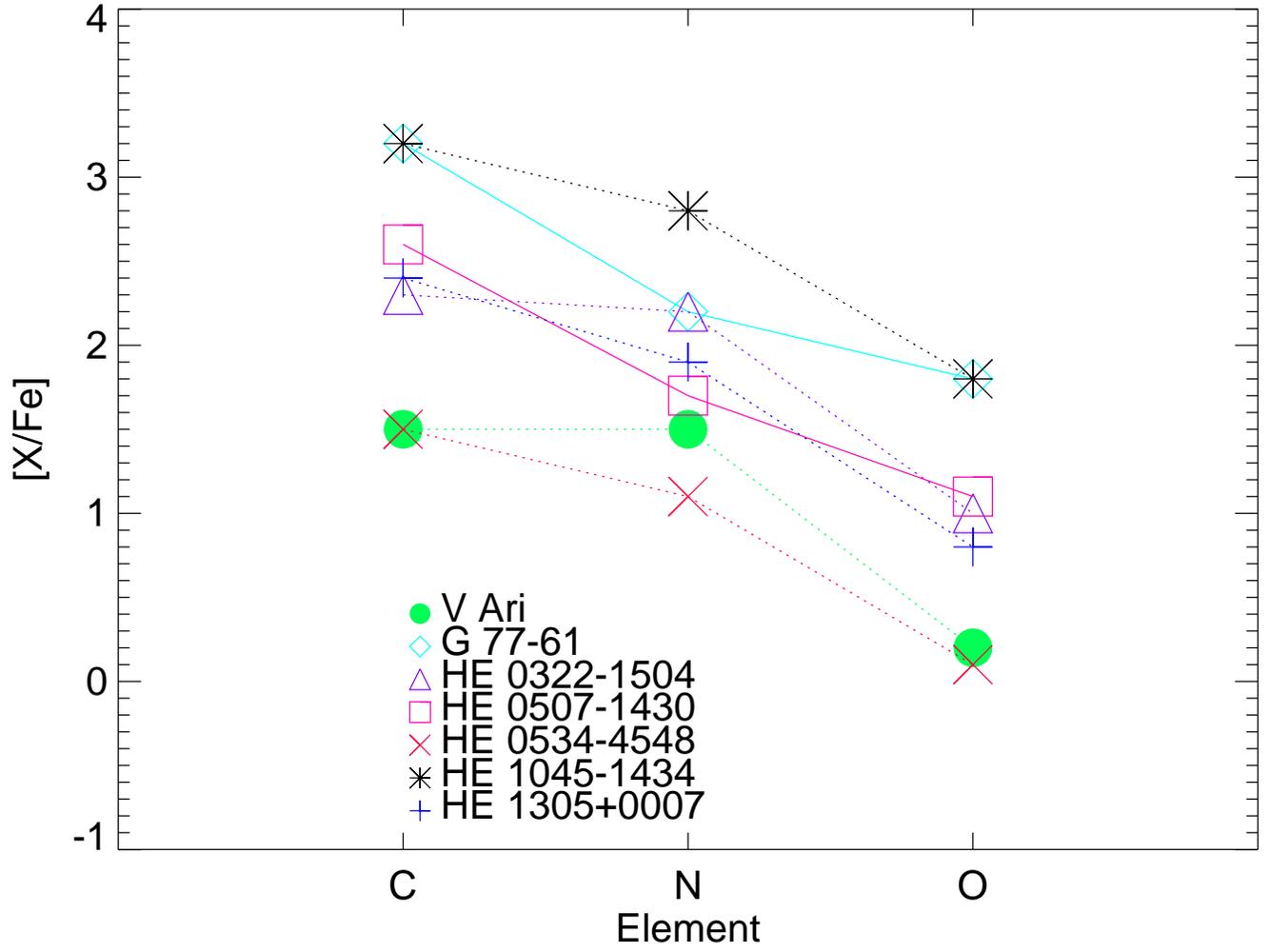}}}
\caption{CNO abundances for the complete sample of CEMP stars studied in this work. }
\end{figure}
\clearpage

\end{document}